\documentstyle[epsfig]{aipproc}

\begin{document}

\newcommand{\nub}{\mbox{$\overline{\nu}$}}
\newcommand{\num}{\mbox{$\nu_\mu$}}
\newcommand{\numb}{\mbox{$\overline{\nu_\mu}$}}
\newcommand{\nue}{\mbox{$\nu_e$}}
\newcommand{\nueb}{\mbox{$\overline{\nu_e}$}}

\title{Detectors for Neutrino Physics at the First Muon Collider}

\author{Deborah A. Harris$^*$ and Kevin S. McFarland$^{\dagger}$}
\address{$^*$University of Rochester, Rochester, NY 14627 \\
 $^{\dagger}$Massachusetts Institute of Technology, 77 Massachusetts Ave., 
Cambridge, MA 02139} 

\maketitle
\vspace{-.25in} 
\begin{abstract} 
We consider possible detector designs for short-baseline neutrino
experiments using neutrino beams produced at the First Muon Collider
complex.  The high fluxes available at the muon collider make possible
high statistics deep-inelastic scattering neutrino experiments with a
low-mass target.  A design of a low-energy neutrino oscillation
experiment on the ``tabletop'' scale is also discussed.
\end{abstract}
\mbox{\parbox{\textwidth}{
\vspace{-8.in} 
\begin{flushright}
LNS-98-276\\ UR-1515 ER$/$4056$/$910 \\ FNAL-CONF-98/114
\end{flushright} 
}}
\vspace{-.75in} 
\section*{Introduction}  

This contribution considers the problem of constructing detectors
appropriate for doing short-baseline neutrino physics at the First
Muon Collider complex.  The physics motivations for these detectors
are discussed elsewhere in these proceedings\cite{phys-proc}.  Since
the proposed experiments are short-baseline, the physics being
considered is primarily the high-energy physics of neutrino-nucleon
deep-inelastic scattering; however, the final section of the paper
considers an oscillation experiment possible with the lowest energy
neutrino beam.

\section*{Neutrino Beams at the Muon Collider Complex} 

The muon collider is expected to use a series of recirculating linacs
to accelerate the muons before injection into a collider ring.  Any
segment along the muon's trajectory that is straight will necessarily
create a collimated neutrino beam with an angular divergence of
approximately $1/\gamma_\mu$.  The recirculating linacs (RLAs) will
have ${\cal O}(300{\rm m})$ in which acceleration takes place, and the
interaction points in the collider rings will also have $5$-$10$m straight
sections. 

We wrote a simple Monte Carlo of the muon collider straight sections
to predict fluxes, based on the workshop muon accelerator parameters.
We used the $\mu^-$ beam, and assumed the beam polarization and
divergence were zero.  Fluxes were predicted far downstream of each
straight section in order to allow for shielding.  We considered only
a ``dumb'' shield option in which sufficient concrete shielding to
range out the muons from the primary muon beam was used.  Table
\ref{tab:flux} gives a table of the distances of a proposed experiment
downstream of each straight section in the RLAs and collider.
\begin{table}[tbp] 
\begin{tabular}{lcccccc} 
Experiment: & RLA1 &  RLA2 & RLA3 & Med Eng & Top & High Eng\\ 
\hline
Max Muon Energy (GeV) & 10 & 70 & 250  & 100 & 175 & 250 \\
Distance  (m) &  23.3 & 163 & 582  & 233 & 408 & 582 \\ 
\hline
 50\% $\nu$'s within radius & $25$~cm & $30$~cm & $25$~cm & 
$15$~cm & $15$~cm & $15$~cm \\
 90\% $\nu$'s within radius & $75$~cm & $75$~cm & $60$~cm & 
$40$~cm & $40$~cm & $40$~cm \\
\hline
 $\nu$ DIS events,($\times 10^6$ ), $r<10cm$  & n/a & 0.038 & 0.11 & 0.13 & 0.21 & 0.34 \\  
 $\nu$ DIS events,($\times 10^6$ ), $r<150cm$ & n/a & 0.49 & 0.92 & 0.54 & 0.99 & 1.34 \\ 
\end{tabular}  
\caption{For each straight section in the RLAs and for several different
collider scenarios, the maximum muon energy, the required shielding distance,
the $\nu$ beam size and the $\nu$ DIS events in Millions per g/cm$^2$/yr are shown.}
\label{tab:flux} 
\end{table} 

\subsection*{Predicted Neutrino Fluxes} 

\begin{figure}[tpb] 
\epsfxsize=\textwidth\epsfysize=3.in
\epsfbox{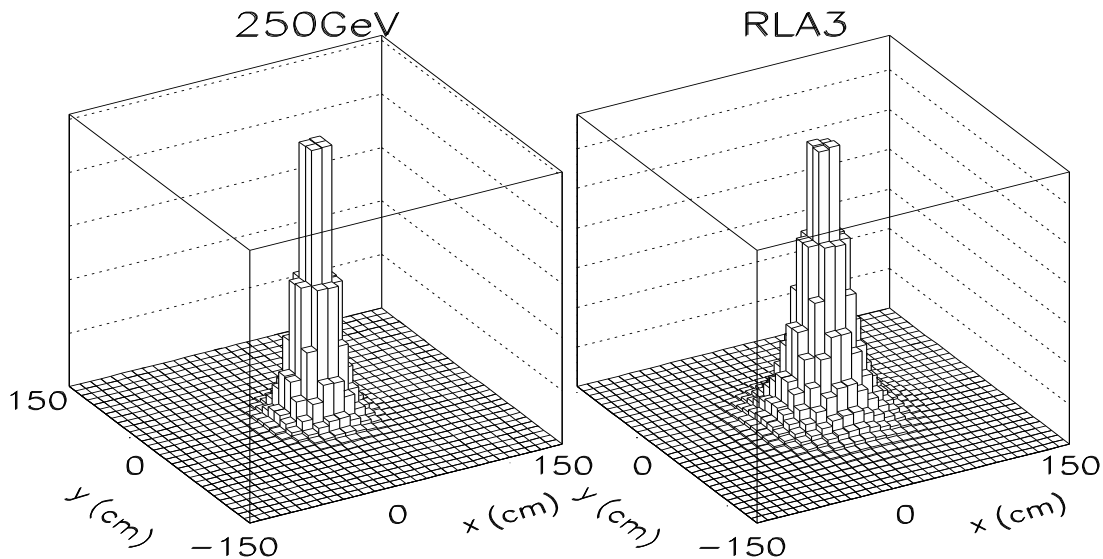} 
\vspace{10pt} 
\caption{Event illuminations in two possible neutrino experiments} 
\label{fig:xyplot} 
\end{figure} 
As shown in Table~\ref{tab:flux}, the beam is relatively small at all
energies (since we have chosen a baseline which is proportional 
to energy due to shielding concerns).  
Figure \ref{fig:xyplot} shows a typical event
illumination for two possible neutrino experiments: one downstream of
RLA3, and one downstream of the 250GeV collider ring's straight
section.


\begin{figure}[tpb] 
\epsfxsize=\textwidth\epsfysize=3.in
\epsfbox{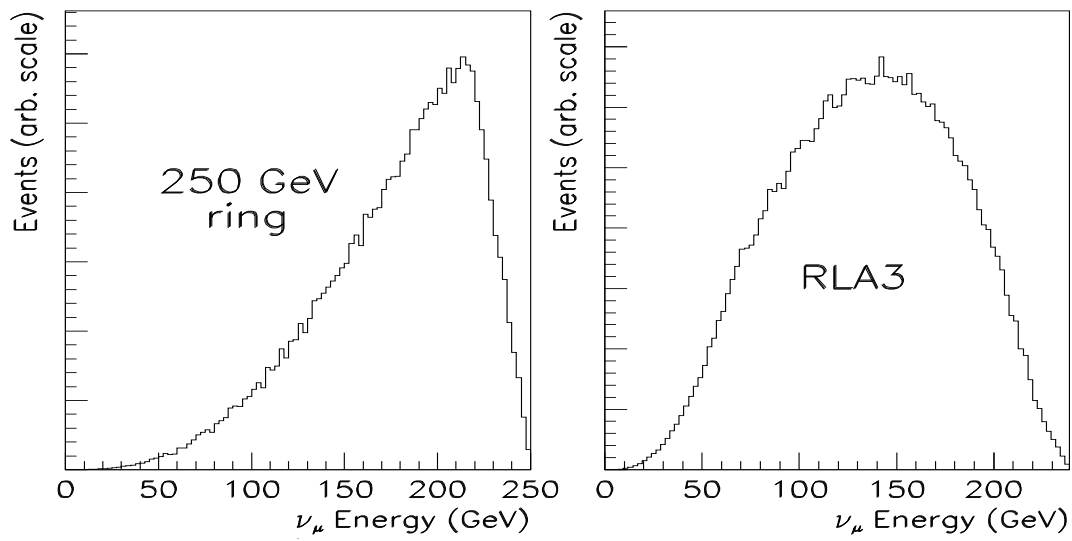} 
\vspace{10pt} 
\caption{Neutrino Energy Spectra for two experiments, each with a 
target of 10 cm radius:  one downstream 
of the 250 GeV collider ring straight section and one downstream of RLA3. 
The broader distribution from RLA3 is an artifact of the energy ramp
in RLA3; turn-by-turn, the spectra have a similar shape.}
\label{fig:energies} 
\end{figure} 
Figure~\ref{fig:energies} shows the neutrino energy spectra reaching a
$10cm$ radius neutrino target downstream of RLA3, and also for the same target
downstream of a 250 GeV collider ring straight section.  Unlike
neutrino beams made by decays of hadrons, the energy spectra will be
well known from the accelerator parameters at a muon collider.  This
will make the difficulties of separating flux and cross-section
relatively trivial and the flux endpoint will provide an important
detector calibration.
 
\subsection*{Muon Flux at Neutrino Experiments} 

\begin{figure}[tbp] 
\epsfxsize=\textwidth\epsfysize=3.in
\epsfbox{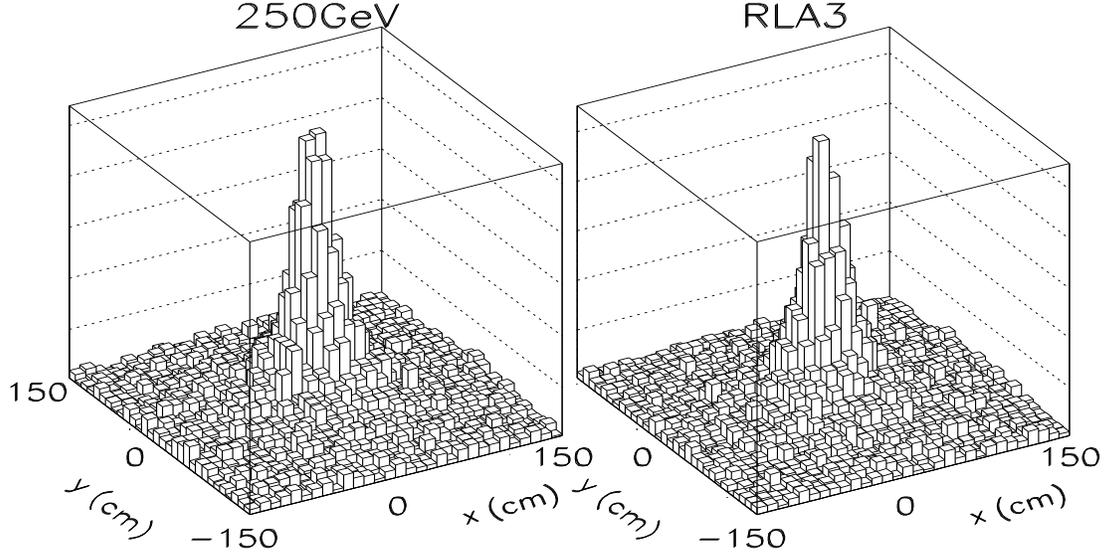} 
\vspace{10pt} 
\caption{Maximum background muon illuminations in one turn for 
two possible neutrino experiments} 
\label{fig:xymuon} 
\end{figure} 
The shielding necessary to protect a neutrino experiment from primary
muons will produce large numbers of muon neutrino DIS interactions
some of which will produce muons in the neutrino detector.  The Monte
Carlo simulation described earlier contained muon production and
propagation in shielding to predict the size of this background.
Figure~\ref{fig:xymuon} shows the surprisingly peaked muon
illumination at the detector for experiments downstream of RLA3 and
the 250 GeV collider using these assumptions.
 
The critical number for an experiment is the maximum number of muons
arriving at the detector for a single turn in the accelerator. This
flux is listed in Table~\ref{tab:backg-pp}.  The background for a detector
measuring $100cm^2$ is always $\leq 0.06$ muons$/$turn and is
therefore not a problem.  However, for large detectors downstream of
the recirculating linacs more complicated shielding would be
necessary.  One option would be to have magnetized shielding followed
by an empty volume which could bend any produced muons away from the
neutrino target.  Another possibility could be to use shielding heavier than
concrete again with a long empty volume just upstream of a large
detector to allow the muons to scatter away from the detector
acceptance.

\begin{table}[tbp]
\begin{tabular}{lcccccc} 
Experiment: &  RLA1 & RLA2 & RLA3 & Med Eng & Top & High Eng\\ 
\hline
Muons/Turn & $0.015$ & $0.06$ & $0.4$ & $0.0015$ & $0.0025$ & $0.0035$ \\
Muons in center $100cm^2$ & 0.0028 & 0.017 & 0.081 & 0.00014 & 0.00024 &0.00033\\
$\nu_\mu$ DIS  events/Turn & $0.05$ & $0.07$ & $0.27$ & $0.0015$ & $0.0015$ & $0.002$ 
\end{tabular}
\caption{Per pulse interaction rates in a $40$~cm$\times40$~cm $10~\lambda_0$ 
steel sampling calorimeter and muon filter located downstream of a ``dumb''
muon filter which ranges out the direct beam muons.}
\label{tab:backg-pp}
\end{table}
 
\section*{High Mass Neutrino Target}  

The fluxes detailed above represent improvements by $3$
to $5$ orders of magnitude over contemporary high energy neutrino
beams\footnote{The NuTeV SSQT had an interaction rate of
${\cal O}(10)$~events/kg/yr with a mean energy of
$120$~GeV\cite{NuTeV}; beams at the CERN PS used by NOMAD and CHORUS
had a mean energy of $35$~GeV and observed
${\cal O}(100)$~events/kg/yr\cite{NOMAD-CHORUS}.}.  It would therefore
be possible to run an experiment with conventional high-mass sampling
target-calorimeters and observe perhaps $10^{11}$ events per year in
previously explored energy regimes.

\begin{figure}[tbp]
\epsfxsize=\textwidth\epsfbox{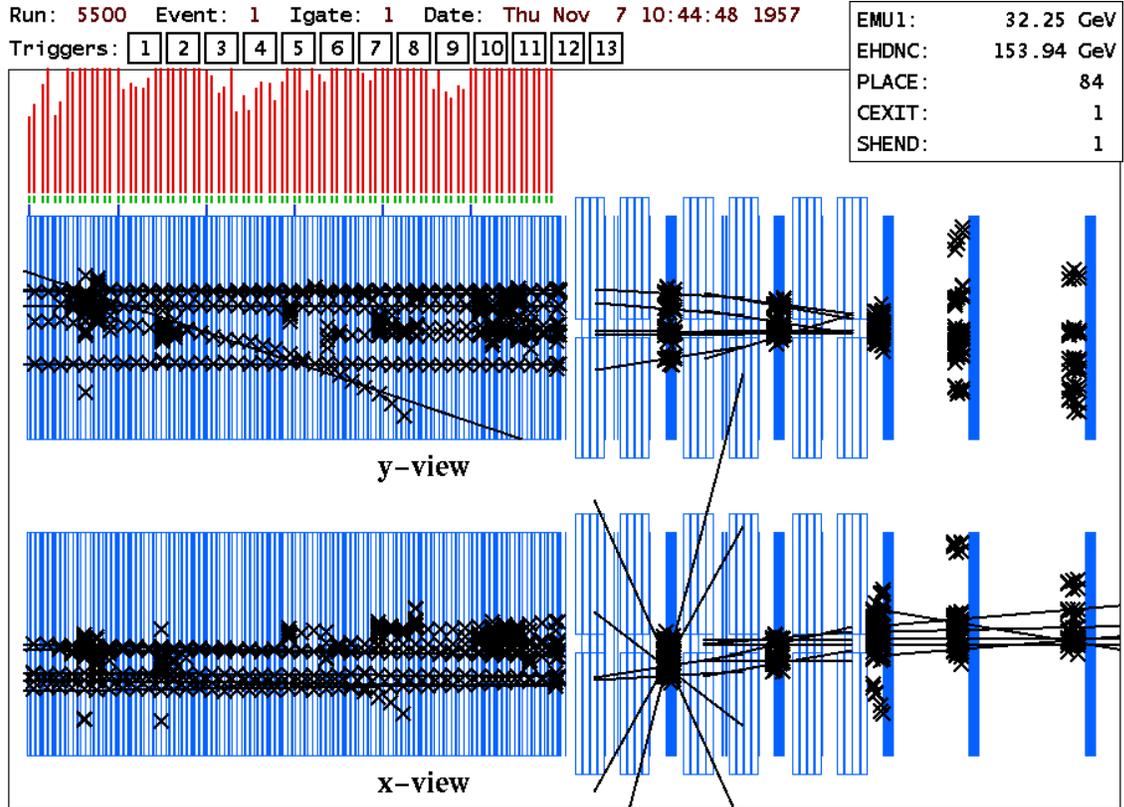}
\caption{An extreme Monte Carlo event in the NuTeV
detector, placed $600$ meters downstream of RLA3.  This event contains
six neutrino interactions and three muons produced in upstream
shielding from a single turn of the muon beam.  This number of muons
would be very unlikely, but the number of neutrino interactions is
below average in the NuTeV detector}
\label{fig:nutev-pileup}
\end{figure}

However, sampling calorimeters would have some difficulties in the FMC
environment.  High mass sampling calorimeters are only designed for
the detection of \num\ or \numb\ since \nue\ and \nueb\ events are
difficult to separate from neutral current interactions of other
neutrino species, especially when the final state lepton has a low 
fraction of the initial neutrino energy.  Therefore, unless the muon
beam polarity of the FMC were reversible, only neutrinos or
anti-neutrinos could be observed in a single detector.  Furthermore,
high mass, sparsely instrumented detectors are sensitive to pile-up.
As per pulse event rates in Table~\ref{tab:backg-pp}  show, this is a
particular concern downstream of the high-energy recirculating linacs
where the beam is concentrated in time in a relatively small number of
turns.  Figure~\ref{fig:nutev-pileup} shows an extreme illustration of
the difficulties of dealing with pile-up in a sample-calorimeter.  

Assuming these difficulties could somehow be overcome, it is also
unclear what the novel physics available from $10^{11}$ neutrino
interactions in dense material would be.  In general, precision
measurements from neutrinos are systematics limited in such
detectors\footnote{For example, the measurements of
$\alpha_S$\cite{bill-alphas} and $\sin^2\theta_W$\cite{CCFR-final-wma}
in the CCFR experiment} and searches for rare phenomena are background
limited.

\section*{Conventional Fixed-Target Geometry} 

\begin{figure}[tbp]
\hspace*{0.25\textwidth}\epsfxsize=0.5\textwidth\epsfbox{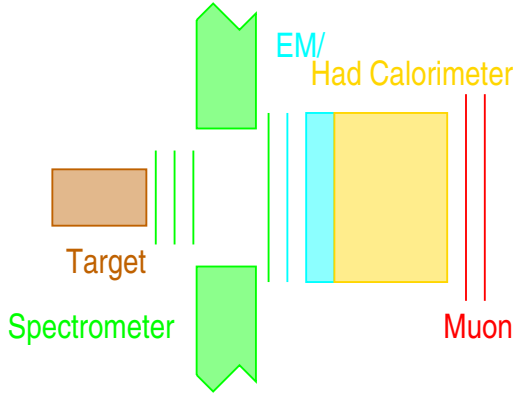}
\caption{A schematic of an FMC neutrino detector in the style of conventional
fixed target detectors.}
\label{fig:ft-schematic}
\end{figure}

Probably more interesting than the above detector technology would be
one where the target is small enough to vary its composition for
studies of nuclear effects on nucleon structure or even spin physics
in neutrino deep inelastic scattering.  This is possible in a geometry
more closely resembling traditional fixed-target experiments: target,
followed by spectrometry and calorimetry and muon identification as
shown in Figure~\ref{fig:ft-schematic}.  The problems of tracking,
particle-ID and calorimetry in such detectors are well-known and will
not be discussed here.  Such a detector would presumably be able to
identify the outgoing lepton in charged-current events, possibly with
TRDs in the downstream tracking providing enhanced identification.
This detector would be able to tag the production of charmed particles
by observing high momentum final state leptons, and with a sufficiently
fine-grained upstream tracker, charmed final state particles could be
tagged {\em via} detached vertices\cite{bjk}.  Physics goals of an
experiment using this detector are described elsewhere in these 
proceedings\cite{phys-proc}.

The most serious technical difficulties of such detectors is pile-up
in the dense calorimetry and muon systems.  A typical
iron-scintillator sampling calorimeter/muon shield of $10~\lambda_0$
($1.6$~m of steel or $13$~kg/cm$^2$) would observe the per pulse
background rates shown in Table~\ref{tab:backg-pp}.  Background rates
at the collider are probably manageable, but the background rates at
the RLAs, particularly RLA3, could be difficult.

\section*{A Table-Top Low Energy Neutrino Detector} 

\begin{figure}[tbp]
\epsfxsize=\textwidth\epsfbox{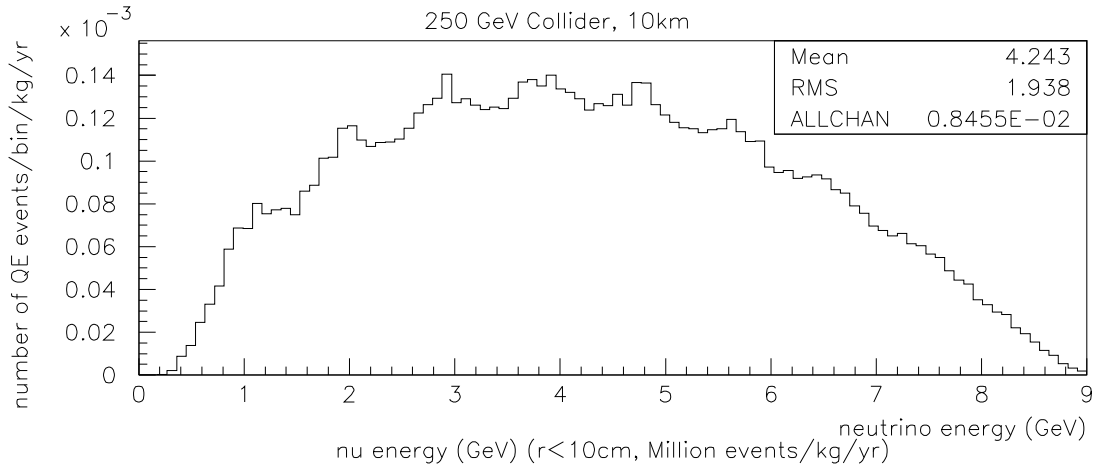}
\caption{Energy spectrum of quasi-elastic $\nu_\mu$ events
downstream of RLA1.}
\label{fig:rla1-qe}
\end{figure}
\begin{figure}[tbp]
\epsfxsize=\textwidth\epsfbox{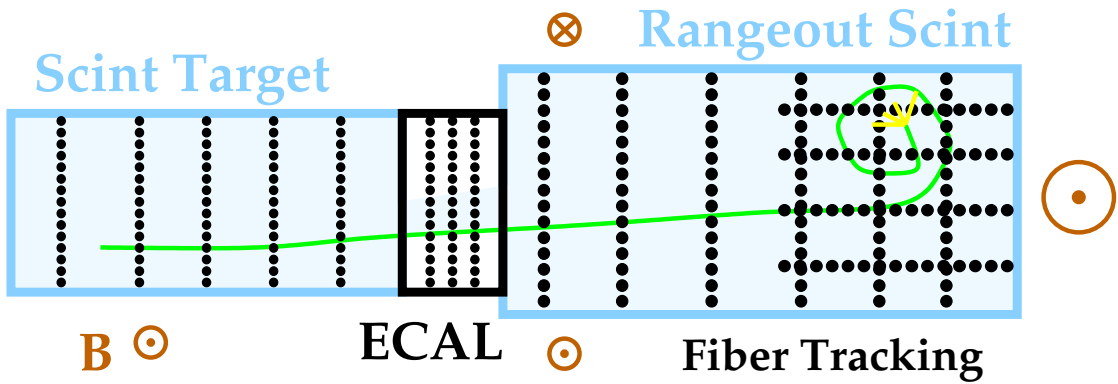}
\caption{A table-top sized neutrino detector for 
that beam.}
\label{fig:table-top-det}
\end{figure}
Another avenue that becomes possible with the FMC neutrino beams is
searching for neutrino oscillations in a modest-sized experiment.
Figure~\ref{fig:rla1-qe} shows the expected energy spectrum of quasi-elastic
\num\ events downstream of RLA1.  In a water-density 0.05m$^3$ target,
one would expect to observe approximately $0.4$~Million events per year.
This sets the scale for a truly ``table-top'' scale neutrino oscillation
experiment.

With a flavor selected beam of \numb\ and \nue, the simplest search
experimentally\footnote{$e^+$ appearance is background-laden at these
energies and $\tau$ appearance, aside from being heavily suppressed by
$m_\tau$, requires a more sophisticated detector.}  is for wrong-sign
muon ($\mu^-$) appearance.  The primary backgrounds would come from
mis-identifying the muon charge or from production of charged pions
which could decay or be misidentified as muons.  By conclusively
identifying $\mu$ charge in a spectrometer, tracking the particle from
the point of appearance and observing its decay with a characteristic
muon lifetime, these backgrounds could presumably be kept very low.  A
naive version of a detector designed for this measurement appears in
Figure~\ref{fig:table-top-det}.  With proper granuality of tracking
and careful optimization of the magnetic field for a broad acceptance
in momentum, even such a small scale detector could reach $10^{-5}$
sensitivity if naive background estimations are correct.

\section*{Conclusions} 

We have presented outlines of detector designs for short-baseline
neutrino experiments at the first muon collider complex.  This facility
would make available neutrino beams of unprecedented intensity, and we argue
that the best way to take advantage of this is to design novel detectors,
such as small targets or ``table-top'' neutrino experiments to explore
different aspects of neutrino scattering or neutrino oscillations
than those studied in the past.

\end{document}